\documentclass[prc,aps,manuscript]{revtex4}
\usepackage{graphicx}
\usepackage{dcolumn}
\usepackage{bm}

\newcommand{\bs}{\bigskip}
\newcommand{\bc}{\begin{center}}
\newcommand{\ec}{\end{center}}

\begin{document}

\title{The effect of the Landau--Zener transitions on nuclear fission dynamics}
\author{\underline{ V. M. \surname{Kolomietz}} \thanks{%
Electronic address: vkolom@kinr.kiev.ua}}
\author{ S. V. \surname{Radionov} \thanks{%
Electronic address: sergey.radionov18@gmail.com}}
\affiliation{\textit{Institute for Nuclear Research, 03680 Kiev, Ukraine}}

\begin{abstract}
In the paper, it is studied the influence of Landau--Zener transitions
between nuclear many--body states on the dissipative properties of nuclear
large--amplitude collective motion. Within the cranking--like approach, we
describe the time evolution of a nuclear many--body system as the
self--consistent motion in a space of intrinsic
excitations and in a space of a single collective (deformation)
parameter. By that we measure how the spectral statistics
of the nuclear energy levels affects the fission rate
at quite large initial temperatures of heavy nuclei.
\end{abstract}

\maketitle



\section{Introduction}

It has been found that the perturbative dynamics of complex quantum systems
has a diffusive behaviour in a space of the occupancies of adiabatic eigenstates
of the system's Hamilton operator. As a result of such a quantum mechanical
diffusion, the complex quantum systems may absorb an energy of external perturbation.
From this perspective, it will be interesting to
measure how the quantum diffusion reveals itself on macroscopic
level, i.e., when the time variations of macroscopic parameter
are not arbitrary but provide a constancy of total energy of the system.
Such a study is important to understand the quantum nature of fluctuations
and dissipation appearing in dynamics of macroscopic (collective)
coordinates in finite Fermi systems.

The main purpose of the present paper is to investigate different time
regimes of the quantum mechanical diffusion and to find its effect on
time evolution of the macroscopic coordinate.
The plan of the paper is as follows. In Sect.~\ref{qdd}, we start from the
time--dependent Schr\"{o}dinger equation and introduce adiabatic basis of
the system's Hamiltonian. In the weak--coupling limit, we get a closed set
of equations for the time evolutions of the occupancies of adiabatic states.
In Sect.~\ref{ca}, we derive equation of motion for the macroscopic
variable. In Sect.~\ref{fr}, we apply the obtained Langevin--like equation
for the macroscopic variable to the problem of thermal overcoming of the
nuclear fission barrier and calculate the fission rate over a model
parabolic barrier. Conclusions and discussion of the main results of
the paper are given in the Summary.

\section{Intrinsic excitations dynamics}

\label{qdd}

We start our discussion of the time evolution of a nuclear many--body
system from the Liouville
equation for the density matrix operator $\hat{\rho}$,
\begin{equation}
\frac{\partial \hat{\rho}(t)}{\partial t}+i\hat{L}(t)\hat{\rho}(t)=0.
\label{L}
\end{equation}%
Here, $\hat{L}$ is the Liouville operator defined in terms of the commutator
\begin{equation}
\hat{L}\hat{\rho}=\frac{1}{\hbar }\left[ \hat{H},\hat{\rho}\right] ,
\label{H}
\end{equation}
where $\hat{H}(t)\equiv \hat{H}[q(t)]$ is a Hamilton operator of the
nuclear system that parametrically depends on a single deformation
(collective) variable $q$.

In the sequel, we shall deal with dynamics of
the density matrix whose diagonal part describes real
transitions between states of the quantum system and which is important
to understand microscopically the appearence of dissipation in macroscopic
collective motion. With that purpose, we apply the Zwanzig's projection
technique~\cite{zw} to Eq.~(\ref{L}),

we introduce a
projection operator $\hat{\mathcal{P}}$ and split the density matrix
operator into the diagonal and non--diagonal parts,
\begin{equation}
\hat{\rho}=\hat{\rho}_{d}+\hat{\rho}_{od},  \label{r12}
\end{equation}
where the diagonal part is defined as
\begin{equation}
\hat{\rho}_{d}=\hat{\mathcal{P}}\hat{\rho},  \label{r1}
\end{equation}
and the non--diagonal part is given by
\begin{equation}
\hat{\rho}_{od}=(1-\hat{\mathcal{P}})\hat{\rho}.  \label{r2}
\end{equation}
It is assumed that the projection operator $\hat{\mathcal{P}}$ is linear and
time--independent. Acting on the Liouville equation~(\ref{L}) by the
operators $\hat{\mathcal{P}}$ and $1-\hat{\mathcal{P}}$ from the left, we
obtain a system of equations for $\hat{\rho}_{d}$ and $\hat{\rho}_{od}$
\begin{equation}
\frac{\partial \hat{\rho}_{d}}{\partial t}+i\hat{\mathcal{P}}\hat{L} (\hat{%
\rho}_{d}+\hat{\rho}_{od})=0,  \label{r1eq}
\end{equation}
\begin{equation}
\frac{\partial \hat{\rho}_{od}}{\partial t}+i(1-\hat{\mathcal{P}})\hat{L} (%
\hat{\rho}_{d}+\hat{\rho}_{od})=0.  \label{r2eq}
\end{equation}

Formally, a solution to Eq.~(\ref{r2eq}) can be written as
\begin{equation}
\hat{\rho}_{od}(t)=\hat{\rho}_{od}(t=0)-i\int_{0}^{t}{\rm exp}\left\{
-i\int_{0}^{t-t^{\prime }}(1-\hat{\mathcal{P}})\hat{L}(t^{\prime \prime
})dt^{\prime \prime }\right\} (1-\hat{\mathcal{P}})\hat{L}(t^{\prime })\hat{%
\rho}_{d}(t^{\prime })dt^{\prime }.  \label{r2sod}
\end{equation}%
Substituting solution~(\ref{r2sod}) into Eq.~(\ref{r1eq}), we obtain a
closed kinetic equation for the diagonal part of the density matrix
operator $\hat{\rho}_{d}$
\begin{eqnarray}
\frac{\partial \hat{\rho}_{d}(t)}{\partial t} &=&-i\hat{\mathcal{P}}\hat{L}%
(t)\hat{\rho}_{d}(t)-i\hat{\mathcal{P}}\hat{L}(t)\hat{\rho}_{od}(t=0)
\nonumber \\
&&+\int_{0}^{t}\hat{\mathcal{P}}\hat{L}(t)\exp \left\{
-i\int_{0}^{t-t^{\prime }}(1-\hat{P})\hat{L}dt^{\prime \prime }\right\} (1-%
\hat{\mathcal{P}})\hat{L}(t^{\prime })\hat{\rho}_{d}(t^{\prime })dt^{\prime
}.  \label{r1s}
\end{eqnarray}

Let us write the basic kinetic equation~(\ref{r1s}) in matrix form. With
that, we use an adiabatic basis of the Hamilton operator $\hat{H}(q)$,
\begin{equation}
\hat{H}(q)\Psi _{n}(q)=E_{n}(q)\Psi _{n}(q).  \label{adiabbas}
\end{equation}%
This basis is determined by a set of static wave functions $\Psi_n$ and
energies $E_{n}$ found at each fixed value of the macroscopic
variable $q$. Using the adiabatic basis~(\ref{adiabbas}), the
time--dependence of the matrix elements of the density matrix operator $\hat{%
\rho}$ are given by
\begin{equation}
\rho _{nm}(t)=\exp \left\{ -i\int_{0}^{t}\omega _{nm}(t^{\prime })dt^{\prime
}\right\} \langle \Psi _{n}|\hat{\rho}|\Psi _{m}\rangle ,  \label{rnm}
\end{equation}%
and the matrix elements of the Liouville operator $\hat{L}$ (\ref{H}) are
equal to
\begin{eqnarray}
(\hat{L})_{nmn'm'}={\rm exp}\left\{
-i\int_{0}^{t}\omega _{nn'}(t')dt'\right\}
\langle \Psi _{n}|\frac{\partial }{\partial t}|\Psi _{n'}\rangle \delta _{mm'}
\nonumber \\
-{\rm exp }\left\{ -i\int_{0}^{t}\omega _{m'm}(t')dt'\right\}
\langle \Psi _{m'}|\frac{\partial }{\partial t}|\Psi _{m}\rangle \delta _{nn'},
\label{Lnm}
\end{eqnarray}%
where $\omega _{nm}=(E_{n}-E_{m})/\hbar $.

The second term on the left--hand side of Eq.~(\ref{r1s}) with the choice of
the projection operator $\hat{\mathcal{P}}$,
\begin{equation}
(\hat{\mathcal{P}})_{nmn^{\prime }m^{\prime }}=\delta _{nm}\delta
_{n^{\prime }m^{\prime }}\delta _{nn^{\prime }},  \label{Pmnmn}
\end{equation}%
vanishes, since in this case
\begin{equation}
\hat{\mathcal{P}}\hat{L}\hat{\mathcal{P}}=0.  \label{PLP}
\end{equation}%
Therefore, from Eq.~(\ref{r1s}) we obtain
\begin{equation}
\frac{\partial \hat{\rho}_{d}(t)}{\partial t}=-\int_{0}^{t}\hat{\mathcal{P}}%
\hat{L}(t)\exp \left\{ -i\int_{0}^{t-t^{\prime }}(1-\hat{P})\hat{L}%
dt^{\prime \prime }\right\} (1-\hat{\mathcal{P}})\hat{L}(t^{\prime })\hat{%
\rho}_{d}(t^{\prime })dt^{\prime }.  \label{r2s}
\end{equation}%
Writing down the last equation in the matrix notations, we obtain
\begin{equation}
\frac{\partial \rho _{nn}(t)}{\partial t}=\sum_{m\neq n}\int_{0}^{t}\mathcal{%
H}_{nnmm}(t,t')[\rho _{mm}(t')-\rho _{nn}(t')]dt'.  \label{kineq}
\end{equation}%
Here, the integral kernel ${\cal H}_{nnmm}$ is equal to
\begin{equation}
\mathcal{H}_{nnmm}(t,t')=-\left( \hat{\mathcal{P}}\hat{L}(t)\exp \left\{
-i\int_{0}^{t-t'}(1-\hat{\mathcal{P}})\hat{L}(t^{\prime })dt^{\prime
}\right\} (1-\hat{\mathcal{P}})\hat{L}(t')\right) _{nnmm},  \label{Hnnmm}
\end{equation}%
where it was used the fact that
\begin{equation}
\sum_{m}\mathcal{H}_{nnmm}=0.  \label{nnnn}
\end{equation}%
We will proceed by considering the integral kernel ${\cal H}_{nnmm}$ of
our basic kinetic equation~(\ref{kineq}). Since the expression (\ref{Lnm})
contains the strongly oscillating exponential factors, we can approximately
put the energy distances $E_{n}-E_{m}$ at the same time instant $t$.
Thus, by using Eq.~(\ref{adiabbas}) one gets from
(\ref{Hnnmm})
\begin{eqnarray}
\mathcal{H}_{nnmm}(t,t') \approx  \sum_{abcd}\frac{\dot{q}(t)\dot{q}(t')}
{(E_{a}-E_{b})(q[t])(E_{c}-E_{d})(q[t])}\left( \frac{\partial
\hat{H}}{\partial q}\right) _{ab}(q[t])\left( \frac{\partial \hat{H}}{%
\partial q}\right) _{cd}^{\ast }(q[t^{\prime }])  \nonumber \\
\times\exp (-i\omega _{ab}(q[t])\cdot t)\exp (i\omega _{cd}(q[t])\cdot t^{\prime
})G_{abcd}(t,t^{\prime })(\delta _{bn}-\delta _{an})(\delta _{dm}-\delta
_{cm}),\label{Hnnmm1}
\end{eqnarray}
where star stands for the complex conjugation and
\begin{equation}
G_{abcd}(t,t^{\prime })=\left( \exp \left\{ -i(1-\hat{\mathcal{P}}%
)\int_{0}^{t-t^{\prime }}\hat{L}(t^{\prime \prime })dt^{\prime \prime
}\right\} \right) _{abcd}.  \label{Gabcd}
\end{equation}
Factor $G_{abcd}$ determines non--perturbative response of the quantum
system~(\ref{L})--(\ref{H}) to the macroscopic parameter's variations.

In this analysis, complexity of a quantum system is understood as the
absence of any special symmetricies in the system. Such system is expected
to have some
universal statistical properties which can be modelled by random matrix
ensembles~\cite{ge}. Within the random matrix approach~\cite{ge}, we average
the right--hand side of Eq.~(\ref{kineq}) over suitably chosen
statistics of the randomly distributed matrix elements $h_{nm}\equiv ({%
\partial \hat{H}}/{\partial q})_{nm}$ and the energy spacings $E_{n}-E_{m}$.
First, we perform the ensemble averaging over the matrix elements. They are
treated as complex random numbers with the real and the imaginary parts
which are independently Gaussian distributed. The correlation functions
of the real and the imaginary parts of the matrix elements can be written
in the following general form \cite{weiden79,wil95}
\begin{equation}
\overline{h_{nm}(q)h_{n^{\prime }m^{\prime }}^{\ast }(q^{\prime })}
=\delta _{nn^{\prime }}\delta _{mm^{\prime }}\frac{\sigma
_{0}^{2}(q)}{\sqrt{\Omega (E_{n})\Omega (E_{m})}\Gamma }f(|E_{n}-E_{m}|/%
\Gamma )Y(q-q^{\prime }).  \label{HH}
\end{equation}
Here, we take into account both the energy and temporal
correlations of the coupling matrix elements $({\partial \hat{H}}/{\partial q})_{nm}$.
In Eq.~(\ref{HH}), $\Omega$ is the average density of states at given excitation energy,
$f$ is the shape, $\sigma_0^2$ is the strength and $\Gamma$ is the width
of the energy distribution of the ensemble averaged matrix elements.
The correlations of the matrix elements, existing at different
values of the macroscopic parameter, $q$ and $q^{\prime }$, are
measured with the help of the correlation function $Y$.
Since the energy correlations between two different states
$n$ and $m$ drop out with the rise of a distance between them,
it is rather obvious that
$f\rightarrow 0$ with $|E_n-E_m|/\Gamma\rightarrow\infty$ and $f \sim 1$
at $|E_n-E_m|/\Gamma<<1$.

In the sequel, we will only study weak--coupling regime of the
driven dynamics (\ref{kineq})--(\ref{Gabcd}) when
\begin{equation}
G_{abcd}=\delta_{ac}\delta_{bd}.
\label{wcr}
\end{equation}
The applicability of such a regime will be discussed later on in next Section.
Under the condition (\ref{wcr}), we obtain (see Eqs.~(\ref{kineq}) and (\ref{HH}))
\begin{eqnarray}
\frac{\partial \overline{\rho }(E_{n},t)}{\partial t} =
\frac{2\sigma_{0}^{2}\dot{q}(t)}{\sqrt{\Omega (E_{n})}\Gamma }\int_{0}^{t}dt'
\int_{E_{gs}}^{+\infty }dE_{m}\sqrt{\Omega (E_{m})}f(|E_{n}-E_{m}|/\Gamma)
\nonumber\\
\times Y(q[t]-q[t^{\prime }])  \frac{\mathrm{cos}%
([E_{n}-E_{m}][t-t^{\prime }/\hbar ])}{(E_{n}-E_{m})^{2}}[\bar{\rho}%
(E_{m},t^{\prime })-\bar{\rho}(E_{n},t^{\prime })],  \label{kineqme}
\end{eqnarray}%
where $E_{gs}$ is the ground--state energy and the summation over all
descrete states $m$ was replaced by the integration over the corresponding
continuous energy variable $E_{m}$.

The energy spacings part of the ensemble averaging procedure is defined
through the two--level correlation function, $R(\Omega|E_{n}-E_{m}|)$, that
is the probability density to find the state $m$ with energy $E_{m}$ within
the interval $[E_{m},E_{m}+dE_{m}]$ at the average distance $|E_{n}-E_{m}|$
from the given state $n$ with energy $E_{n}$. We give an explicit form of
the function $R$ for Gaussian Orthogonal Ensemble (GOE)~\cite{pm}
\begin{equation}
R_{GOE}(x)=1-\left( \frac{\mathrm{sin}(\pi x)}{\pi x}\right) ^{2}+ \left(
\int_{0}^{1}dy\frac{\mathrm{sin}(\pi xy)}{y}-\frac{\pi }{2}\right) \left(
\frac{\mathrm{cos}(\pi x)}{\pi x}-\frac{\mathrm{sin}(\pi x)} {(\pi x)^{2}}%
\right) ,  \label{RGOE}
\end{equation}

Gaussian Unitary Ensemble (GUE)
\begin{equation}
R_{GUE}(x)=1-\left( \frac{\mathrm{sin}(\pi x)}{\pi x}\right) ^{2},
\label{RGUE}
\end{equation}

and Gaussian Symplectic Ensemble (GSE) of levels
\begin{equation}
R_{GSE}(x)=1-\left( \frac{\mathrm{sin}(2\pi x)}{2\pi x}\right) ^{2}
+\int_{0}^{1}dy\frac{\mathrm{sin}(2\pi xy)}{y}
\left(\frac{{\rm cos}(2\pi x)}{2\pi x}-\frac{{\rm sin}(2\pi x}{(2\pi x)^2}\right),
\label{RGSE}
\end{equation}
where $x\equiv |E_{n}-E_{m}|\Omega (E_{n})$. The behaviour of the two--level
correlation function $R(x)$ with the normalized level spacing $x$ for the
different statistical ensembles~(\ref{RGOE}), (\ref{RGUE}) and (\ref{RGSE})
is shown in
\textrm{Fig.~1}. The main difference between the statistics, seen in
\textrm{Fig.~1}, is the behaviour of $R(x)$ at small energy spacings $x$.
For the GOE statistics one has the linear repulsion between levels, $%
R_{GOE}\sim x$, the GUE statistics implies the quadratic level
repulsion, $R_{GUE}\sim x^{2}$, while in the GSE case we have $R_{GSE}\sim x^{4}$.
On the other hand, $R_{GOE}$, $R_{GUE}$ and $R_{GSE}$
are similar at moderate spacings $x$, when the spectral correlations between
levels consistently disappear.

Introducing the new energy variables,
\begin{equation}
E\equiv E_{n},~~~e\equiv E_{n}-E_{m},  \label{eee}
\end{equation}
we rewrite the dynamical equation (\ref{kineqme}) for the occupancies of the
quantum adiabatic states within the random matrix approach as
\begin{eqnarray}
\frac{\partial \overline{\rho }(E,t)}{\partial t} =\frac{2\sigma^2_0 \dot{q}%
(t)}{\sqrt{\Omega (E)}\Gamma}\int_{0}^{t}dt^{\prime}\dot{q}%
(t^{\prime})Y(q[t]-q[t^{\prime}]) \int_{-\infty }^{+\infty }de\sqrt{\Omega
(E-e)}R(\Omega|e|)f(|e|/\Gamma)
\nonumber \\
\times\frac{\cos (e[t-t^{\prime}]/\hbar )}{e^{2}} [%
\bar{\rho}(E-e,t^{\prime})-\bar{\rho}(E,t^{\prime})].  \label{kineq2}
\end{eqnarray}
The integration limits over the energy spacing $e$ were extended to
infinities since the time changes of the occupancy $\bar{\rho}(E,t)$ of the
given state with the energy $E$ are mainly due to the direct interlevel
transitions from the close--lying states located at the distances $|e|<<E$.
The same assumptions enable us to truncate expansion to $e^{3}$--order
terms,
\begin{eqnarray}
\sqrt{\Omega (E-e)}[\bar{\rho}(E-e,t^{\prime})-\bar{\rho}(E,t^{\prime})]=-%
\sqrt{\Omega (E)} \frac{\partial \bar{\rho}(E,t^{\prime})}{\partial E}e
\nonumber \\
+\frac{1}{2\sqrt{\Omega (E)}}\frac{d\Omega (E)}{dE}\frac{\partial \bar{\rho}%
(E,t^{\prime})}{\partial E}e^{2}+\frac{\sqrt{\Omega (E)}}{2}\frac{\partial
^{2} \bar{\rho}(E,t^{\prime})}{\partial E^{2}}e^{2}+O(e^{3}).  \label{expan}
\end{eqnarray}
Substituting the expansion~(\ref{expan}) into Eq.~(\ref{kineq2}), the odd-$e$
terms drop out and we obtain the diffusion--like equation of motion for the
occupancy $\overline{\rho }(E,t)$,
\begin{equation}
\Omega (E)\frac{\partial \bar{\rho}(E,t)}{\partial t}\approx \sigma^2_0\dot{q%
}(t)\int_{0}^{t}dt^{\prime}K(t,t')\dot{q}(t^{\prime})%
\frac{\partial }{\partial E} \bigg[\Omega (E)\frac{\partial \bar{\rho}%
(E,t^{\prime})}{\partial E}\bigg].  \label{rnonm}
\end{equation}
In Eq.~(\ref{rnonm}), the memory kernel, $K(t,t')$, is
defined as
\begin{equation}
K(t,t')=\frac{1}{\Gamma}Y(q-q^{\prime})\int_{\infty }^{+\infty }de
f(|e|/\Gamma)R(|e|\Omega
(E))\mathrm{cos}(e[t-t^{\prime}]/\hbar).  \label{Kts}
\end{equation}
One can treat dynamical process~(\ref{rnonm}) as a quantum mechanical
diffusion of energy in space of the occupancies of quantum adiabatic states,
where $\Omega (E)\bar{\rho}(E,t)$ gives a probability density to find the
quantum system with energy lying in the interval $[E,E+dE]$ at the moment of
time $t$. Due to our truncation of the expansion~(\ref{expan}), the
diffusion is Gaussian. Certainly, keeping the higher order terms in Eq.~(\ref%
{expan}) would imply a non--Gaussian character of the diffusive process~(\ref%
{rnonm}).

\section{Different regimes of the quantum mechanical diffusion}

\label{drqdd}

Features of the perturbed dynamics~(\ref{rnonm}) are defined by the static
characteristics of the quantum system, its average density of states $\Omega
$, the strength $\sigma _{0}^{2}$ and width $\Gamma $ of the energy
distribution~(\ref{HH}) of the coupling matrix elements~(\ref{HH}), as well
as by the amplitude $\Delta q\equiv q(t)-q(t=0)$ and velocity $\dot{q}(t)$
of the macroscopic coordinate's variations. First, we shall investigate
under what conditions the weak--coupling regime (\ref{wcr}) of the
driven dynamics is realized. With this purpose, we ensemble average the
non--perturbative factor $G_{abcd}$ (\ref{Gabcd}). One sees that
$\overline{G}_{abcd}$ will only contain even powers of the
Liouville operator $\hat{L}$ and such an expansion in terms of $\hat{L}$
is determined by a perturbation parameter
\begin{equation}
\alpha \sim \bigg|\bigg(\int_{0}^{t}dt_{1}\int_{0}^{t_{1}}dt_{2}\overline{(1-%
\hat{\mathcal{P}})\hat{L}(t_{1})(1-\hat{\mathcal{P}})\hat{L}(t_{2})}\bigg)%
_{abab}\bigg|.  \label{alpha}
\end{equation}%
With the help of Eqs.~(\ref{Pmnmn}) and (\ref{HH}), one can show that
\begin{eqnarray}
\alpha &=&\frac{\sigma _{0}^{2}}{\Gamma }\bigg|\int_{0}^{t}dt_{1}\dot{q}%
(t_{1})\int_{0}^{t_{1}}dt_{2}\dot{q}(t_{2})\bigg\{\sum_{c}\frac{%
f(|E_{c}-E_{a}|/\Gamma)}{(E_{c}-E_{a})^{2}\sqrt{\Omega (E_{c})\Omega (E_{a})}}\mathrm{%
cos}\bigg(\frac{[E_{c}-E_{a}][t_{1}-t_{2}]}{\hbar }\bigg)  \nonumber \\
&&+\sum_{d}\frac{f(|E_{d}-E_{b}|/\Gamma)}{(E_{d}-E_{b})^{2}\sqrt{\Omega (E_{d})\Omega
(E_{m})}}\mathrm{cos}\bigg(\frac{[E_{d}-E_{b}][t_{1}-t_{2}]}{\hbar }\bigg)
\nonumber \\
&&-2\frac{f(|E_{a}-E_{b}|/\Gamma)}{(E_{a}-E_{b})^{2}\sqrt{\Omega (E_{a})\Omega (E_{b})%
}}\mathrm{cos}\bigg(\frac{[E_{a}-E_{b}][t_{1}-t_{2}]}{\hbar }\bigg)\bigg\}%
\bigg|.  \label{alpha1}
\end{eqnarray}%
Making simplifying assumption on equidistant spectrum of the quantum system,
we get a condition for the applicability of the weak--coupling regime (\ref{wcr}):
\begin{equation}
\alpha \sim \frac{\sigma _{0}^{2}\Omega ^{3}(\Delta q)^{2}}{\Omega \Gamma }<<1.
\label{wcral}
\end{equation}

Our next goal here is to discuss time scales defining the
driven dynamics. Three different timescales enter the diffusive dynamics
(\ref{rnonm})--(\ref{Kts}). The first timescale, $\tau_{cor}$, is
characteristic interval in time over which the
coupling matrix elements $({\partial \hat{H}}/{\partial q})_{nm}(q[t])$
and $({\partial \hat{H}}/{\partial q})_{nm}^{*}(q[t'])$ are effectively
correlate.
Putting $q[t']\approx q[t]+\dot{q}t$ and by using perturbation theory with
respect to small parameter $|\dot{q}t/\Delta q|$, we can estimate the
correlation time $\tau_{cor}$ as \cite{wil95}
\begin{equation}
\tau_{cor}\sim \frac{1}{(\sigma_0/\sqrt{\Omega\Gamma})\Omega\dot{q}}.
\label{tcor}
\end{equation}
The second one is the characteristic timescale $\hbar/\Gamma$ caused by
the finite width $\Gamma$ of the energy distribution of the
ensemble averaged matrix elements (\ref{HH}). In the limit
$\Gamma>>1/\Omega$ (when the features of the location
of neighboring energy levels (\ref{RGOE})--(\ref{RGUE}) are unsignificant),
the memory kernel $K(t,t')$ (\ref{Kts}) is defined by the
${\rm cosine}$--Fourier transform of the matrix elements' energy distribution:
\begin{equation}
K(t,t')=2\pi Y(q-q'){\cal F}_{\rm cos}\bigg(f[(t-t')/(\hbar/\Gamma)]\bigg).
\label{KF}
\end{equation}
And the third timescale is the characteristic time of the macroscopic coordinate's
variations $\tau_{macr}$. In the paper, we do not investigate the effect
of the time correlations of the coupling matrix elements by putting
\begin{equation}
Y(q-q')=1,
\label{Y}
\end{equation}
This is so when we neglect the time variations of the coupling matrix
elements,
$(\partial \hat{H}/\partial q)_{nm}(q[t])\approx (\partial
\hat{H}/\partial q)_{nm}(q[t=0])$.
On the other hand, the condition (\ref{Y}) can be considered as a consequence
of the weak--coupling limit (\ref{wcral}) since in this case, the correlation
time $\tau _{cor}$  (\ref{tcor}) of the matrix elements is the largest
timescale in the system,
\begin{equation}
\tau _{cor}>>\tau _{macr}~~~~~~~~~~~~
(\frac{1}{(\sigma _{0}/\sqrt{\Omega\Gamma })\Omega \dot{q}}>>\tau _{macr}).
\label{tcortcoll}
\end{equation}
At the end of this discussion, we would like to point out the following fact.
As is shown in Refs.~\cite{wil91,wil95}, the existence of temporal correlations
between the coupling matrix elements plays an important role at the
non--perturbative regime of parametrically driven dynamics of complex
quantum systems. Thus, taking into account of these correlations gives
rise to significant reduction of the diffusion coefficient compared to the
well--known one given by the Kubo formula \cite{wil95}.

In fact, the different dynamical regimes of the quantum
mechanical diffusion (\ref{rnonm}) is defined by the
relationship between the characteristic time $\tau_{macr}$ of the
macroscopic coordinate's variations and the characteristic
time scale of the driven dynamics $\hbar/\Gamma$:

\bs

(i)~$\hbar/\Gamma <<\tau_{macr}$. To understand
this limiting situation, when the interaction $f$ (\ref{HH}) between the
complex quantum states almost the same over a large energy window $\Gamma $,
we assume at the moment that initially only one fixed eigenstate $n_{0}$ of
the system is occupied. Then, with a time
run, the initial occupation peak will spread out over a huge number of
neighboring states. One can also say that the quantum chaotic system~(\ref{L}%
)--(\ref{H}) adopts almost instantentiously to the external perturbation $%
\dot{q}(t)$. In this case, the memory kernel~(\ref{KF}) is sharply peaked
with respect to $t-t^{\prime }$ and we get a normal diffusive regime of the
quantum driven dynamics~(\ref{rnonm}):
\begin{equation}
\Omega (E)\frac{\partial \bar{\rho}(E,t)}{\partial t}= \frac{\hbar \sigma
_{0}^{2}\dot{q}^{2}(t)}{\Gamma }\frac{\partial }{\partial E}\bigg[\Omega (E)%
\frac{\partial \bar{\rho}(E,t)}{\partial E}\bigg].  \label{dif0}
\end{equation}
Such situation corresponds to the usual Kubo formula regime,
when the energy diffusion coefficient $D_E=\hbar\sigma_0^2\dot{q}^2/\Gamma$
is proportional to
the square of the external parameter's velocity $\dot{q}^2$.
Formally, at $\Gamma =\infty $, the energy diffusion disappears since all
the eigenstates becomes equally occupied implying the absence of any energy
flows in the system and therefore, any kind of diffusion.

\bs

(ii)~$\hbar/\Gamma>>\tau_{macr}$. Now, each
eigenstate is effectively coupled only to a few neighboring states and
therefore, the initial occupation distribution will slightly disperse and
remain almost unchanging for a quite long time of order $\hbar/\Gamma$. The
memory kernel $K(t,t^{\prime})$ can be well approximated by constant at $%
\hbar/\Gamma\rightarrow\infty$ and the non--Markovian diffusion--like
equation~(\ref{rnonm}) for the occupancies becomes a wave equation of the
following form:
\begin{equation}
\Omega (E)\frac{\partial^2 \bar{\rho}(E,t)}{\partial t^2}= \sigma^2_0
\dot{q}(t)\Delta q(t)\frac{\partial }{\partial E} \bigg[\Omega (E)\frac{%
\partial \bar{\rho}(E,t)}{\partial E}\bigg].  \label{dif8}
\end{equation}
In fact, the transport of energy from the occupied states to unoccupied ones
undergoes as a wave propagation with the speed $\sigma^2_0\dot{q}\Delta q$.
Here, we have a ballistic regime of the quantum driven dynamics~(\ref{rnonm}).

\bs

(iii)~At moderate values of the width $\Gamma$, the quantum driven dynamics~(%
\ref{rnonm}) is essentially influenced by memory effects, when both the
diffusive and ballistic regimes coexist. At the constant driven velocity, $%
\dot{q}(t)=q_0$, in the beginning the ballistic regime of the quantum
mechanical diffusion~(\ref{rnonm}) sets in
\begin{equation}
var_E\equiv \langle E^2 \rangle-\langle E \rangle^2 \approx \sigma^2_0 \dot{q%
}_0^2 \cdot t^2,~~~~~0<t \leq \frac{\hbar}{\Gamma},  \label{bal}
\end{equation}
where
\begin{equation}
\langle ... \rangle \equiv \int_{E_{gs}}^{+\infty} ...~\bar{\rho}(E,t)
\Omega(E)dE.  \label{<>}
\end{equation}
The ballistic energy transport is relieved by the the normal energy
diffusion
\begin{equation}
var_E\approx \frac{\hbar\sigma^2_0 \dot{q}_0^2}{\Gamma} \cdot t,~~~~~t \geq
\frac{\hbar}{\Gamma}.  \label{dif}
\end{equation}

\section{Cranking approach}

\label{ca}

It is rather interesting to see how the dynamical regimes (\ref{dif0}%
)--(\ref{dif8}) of the quantum mechanical diffusion show up macroscopically.
More specifically, would it be a microscopic source for damping of different
types of collective excitations in complex many--body systems, when the
macroscopic collective modes of motion are coupled to infinite bath of the
intrinsic degrees of freedom?  The appropriate
approach for such investigation is a cranking model. Following the ideology
of this approach, we define the time evolution of a macroscopic collective
variable from the condition of the energy conservation.

To treat selfconsistently dynamics of the classical macroscopic coordinate and
quantum system within the cranking approach, one has to
clarify the following. The energy diffusion regime of the perturbed dynamics
implies statistic interpretation. One can say that actually we have \textit{%
an ensemble} of quantum chaotic system, characterizing by its own initial
density matrix operator $\hat{\rho}(t=0)$ and energy path $E(t)$ (\ref{rnonm}%
). Therefore, in order to provide the constancy of the total energy of the system,
one has necessarily to introduce \textit{a bunch} of
macroscopic parameter's trajectories. Or, we are able to claim that the time
evolution of the macroscopic classical parameter should be \textit{random}
(non--deterministic). By that, we microscopically derived the
fluctuations in the macroscopic collective dynamics.

To obtain an equation of motion for the driving parameter, we first find the
average energy of the system, $\Sigma(t)=Tr[\hat{H}\{q(t)\},\rho(t)]$.
Calculating its time change, we get
\begin{equation}
\frac{d\Sigma }{dt}=\sum_{i}\dot{q}_{i}\frac{\partial E_{\mathrm{gs}}}{%
\partial q_{i}}+\sum_{i}\dot{q}_{i}\sum_{n,m}\left( \frac{\partial \hat{H}}{%
\partial q_{i}}\right) _{mn}\rho _{nm}+\sum_{n}E_{n}\frac{\partial \rho _{nn}%
}{\partial t}+\sum_{i}\dot{q}_{i}\sum_{n}\left( \frac{\partial \hat{H}}{%
\partial q_{i}}\right) _{nn}\rho _{nn}.  \label{dEdt}
\end{equation}

The first term in the right--hand side of Eq.~(\ref{dEdt}) describes a
change of the macroscopic potential energy. The second contribution to the
energy rate $d\Sigma /dt$ is defined by the non--diagonal components of the
density matrix $\rho _{nm}(t)$. Its time evolution is caused by the virtual
transitions among the adiabatic states. We believe that such a term is a
microscopic source for the appearance of the macroscopic kinetic energy. To
demonstrate that, we write it as
\begin{equation}
\left( \frac{d\Sigma }{dt}\right) ^{\mathrm{virt}}\equiv \dot{q}%
\sum_{nm}\left( \frac{\partial \hat{H}}{\partial q}\right) _{mn}\rho _{nm}=2%
\dot{q}\sum_{nm}\int_{0}^{t}dt^{\prime }\ \mathcal{V}_{nm}(t,t^{\prime })%
\dot{q}(t^{\prime })[\rho _{mm}(t^{\prime })-\rho _{nn}(t^{\prime })],
\label{Evt1}
\end{equation}%
where
\begin{equation}
\mathcal{V}_{nm}(t,t^{\prime })=h_{nm}(t)h_{mn}(t^{\prime })\frac{\mathrm{cos%
}(\omega _{nm}[t-t^{\prime }])}{\omega _{nm}}.  \label{V}
\end{equation}%
We formally extend the lower limit of the time integration in Eq.~(\ref{Evt1}%
) to $-\infty $. In this way, we would like to study stationary dynamics of
the complex quantum system, i. e., when the dynamics of the system does not
depend on the choice of initial time.  Thus, integrating by parts the time
integral in the right--hand side of Eq.~(\ref{Evt1}), one can show that
\begin{equation}
\int_{-\infty }^{t}dt^{\prime }\mathcal{V}_{nm}(t,t^{\prime })\dot{q}%
(t^{\prime })[\rho _{mm}(t^{\prime })-\rho _{nn}(t^{\prime })]\approx
\sum_{l=0}^{+\infty }\omega _{nm}^{-(2l+3)}\times \frac{d^{(2l+1)}\left(
\dot{q}h_{nm}h_{mn}[\rho _{mm}-\rho _{nn}]\right) }{dt^{(2l+1)}}.
\label{tparts}
\end{equation}%
In the weak--coupling limit~(\ref{wcr}), we obtain
\begin{equation}
\left( \frac{d\Sigma }{dt}\right) ^{\mathrm{virt}}\approx \dot{q}B(q)\ddot{q}%
+\frac{\partial B}{\partial q}\dot{q}^{3},  \label{dEdtvir}
\end{equation}%
where a term
\begin{equation}
M=\sum_{n,m}h_{nm}h_{mn}\omega _{nm}^{-3}[\rho _{mm}-\rho _{nn}]  \label{B}
\end{equation}%
can be associated with a macroscopic inertia coefficient.

The third term in the right--hand side of Eq.~(\ref{dEdt}) is determined by
the real transitions between the adiabatic states thus, defining how the
energy of macroscopic motion is transfered into the energy of the intrinsic
excitations of the quantum chaotic system:
\begin{equation}
\left( \frac{d\Sigma }{dt}\right)^{\mathrm{real}}= \sigma^2_0\dot{q}%
(t)\int_{0}^{t}dt^{\prime}K(t,t^{\prime})\dot{q}(t^{\prime})
\int_{E_{gs}}^{+\infty}dE\Omega(E)E \frac{\partial }{\partial E}\bigg[%
\Omega(E)\frac{\partial \bar{\rho}(E,t^{\prime})} {\partial E}\bigg]+\dot{q}%
(t)\xi(t),  \label{dEdtreal}
\end{equation}
where Eq.~(\ref{rnonm}) was used. $\xi(t)$ in Eq.~(\ref{dEdtreal}) is a
stochastic term whose ensemble averaged value is zero. By using Eq.~(\ref{HH}%
), one can show that its correlation function is given by
\begin{equation}
\langle\xi(t)\xi(t^{\prime})\rangle=4\sum_{nk} \overline{h_{nk}h^*_{nk}}|%
\overline{\rho}_{nk}(t=0)|^2 \mathrm{cos}[\omega_{nk}(t-t^{\prime})].
\label{xixi}
\end{equation}

The fourth term in the r.h.s of Eq.~(\ref{dEdt}) is given by the
distribution of slopes of the adiabatic eigenstates $E_{nn}$. Within the
random matrix model~(\ref{HH})--(\ref{RGSE}), the negative and positive
slopes of the adiabatic states are assumed to be equally distributed.
Therefore, under the averaging over all random
realizations of the random matrices, modeling the nuclear many body
spectrum, one can neglect the contribution from the fourth term in the rhs
of Eq.~(\ref{dEdt}).

Thus, putting together different contributions (\ref{dEdtvir}) and (\ref%
{dEdtreal}) to the energy rate (\ref{dEdt}), we get transport description of
the macroscopic coordinate's dynamics
\begin{equation}
M\ddot{q}=-\frac{\partial M}{\partial q}\dot{q}^2
-\frac{\partial E_{\mathrm{gs}}}{\partial q}-\sigma
_{0}^{2}\int_{0}^{t}dt^{\prime }K(t,t^{\prime })\dot{q}(t^{\prime
})\int_{E_{gs}}^{+\infty }dE\Omega (E)E\frac{\partial }{\partial E}\bigg[%
\Omega (E)\frac{\partial \bar{\rho}(E,t^{\prime })}{\partial E}\bigg]-\xi
(t).  \label{eqm}
\end{equation}%
The transport equation~(\ref{eqm}) should be supplemented by an equation of
motion for the occupancies of the quantum states (\ref{rnonm}). Importantly
that the dynamics of the external macroscopic parameter is damped only when
the average level--density of the quantum chaotic system $\Omega (E)$ is a
growing function of the intrinsic excitation $E$. If the initial excitation $%
E_{0}$ of the quantum system is sufficiently large than the typical
variations of the energy associated with the macroscopic parameter then, one
can obtain approximately
\begin{equation}
M\ddot{q}=-\frac{\partial M}{\partial q}\dot{q}^2
-\frac{\partial E_{\mathrm{gs}}}{\partial q}-\sigma _{0}^{2}\frac{%
d\Omega (E)/dE}{\Omega (E)}\bigg|_{E_{0}}\int_{0}^{t}dt^{\prime
}K(t,t^{\prime })\dot{q}(t^{\prime })-\xi (t),  \label{eqm1}
\end{equation}%
where the normalization condition
\begin{equation}
\int_{E_{gs}}^{+\infty }\bar{\rho}(E,t)\Omega (E)dE=1  \label{norm}
\end{equation}%
was used.

\section{Nuclear fission rate}
\label{fr}

To measure the role of memory effects in collective dynamics of the
nuclear system on the way from ground state to saddle point, we restrict
ourselves by considering a one--dimensional collective motion $q(t)$
over a schematic parabolic barrier. The potential
energy $E_{\mathrm{pot}}$ presents a single--well barrier formed by a smoothing
joining at $q=q^{\ast }$ of the potential minimum oscillator with the
inverted oscillator
\begin{eqnarray}
E_{\mathrm{pot}}=\frac{1}{2}M\omega _{A}^{2}(q-q_{A})^{2},~~~~q\leq q^{\ast
},  
\nonumber\\
=E_{\mathrm{pot,B}}-\frac{1}{2}M\omega _{B}^{2}(q-q_{B})^{2},~~~~q>q^{\ast}.
\label{Epot}
\end{eqnarray}



We also use a constant value for the mass parameter (\ref{B}),
\begin{equation}
M=\frac{1}{5}AmR_0^2,
\label{Bhydr}
\end{equation}
where $A$ is the mass number of a nucleus, $m$ is the nucleon mass
and $R_0$ is the radius of the nucleus. In numerical calculations,
it was solved the generalized Langevin equation:
\begin{equation}
M\ddot{q}=-\frac{\partial E_{\rm pot}}{\partial q}-
{\kappa_0}\int_0^t e^{-|t-t'|/\tau} \dot{q}(t')dt'-
\xi(t),
\label{langeq}
\end{equation}
where $\tau$ is a memory time and the stochastic force term
$\xi(t)$ in the last equation is related to the memory--dependent
friction force as
\begin{equation}
\langle \xi(t)\xi(t') \rangle = T \kappa_0 e^{-|t-t'|/\tau}.
\label{xixi}
\end{equation}
Here,
\begin{equation}
T=\frac{d\Omega (E)/dE}{\Omega (E)}\bigg|_{E_{0}}
\label{T}
\end{equation}
is understood as an initial temperature of the nucleus.

In Fig.~1, we plotted an escape rate $R_f$ over the parabolic
potential barrier as a function of the initial temperature $T$
for quite small memory time $\tau=2*10^{-23}~s$ (points line)
and for fairly large memory time $\tau=8*10^{-23}~s$ (solid line).

\begin{figure}[h]
\includegraphics[width=70mm,height=70mm, angle=-90]{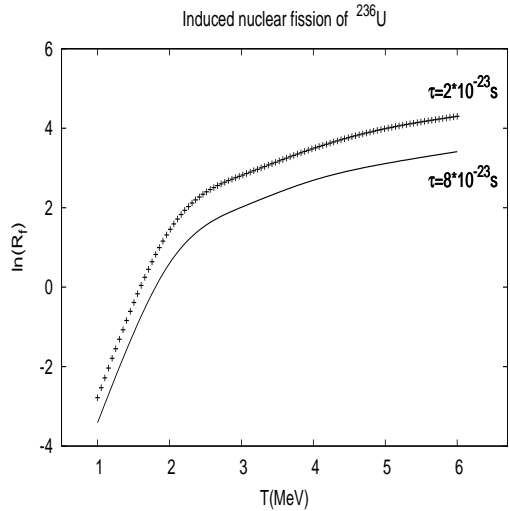}
\caption{The fission rate $R_f$ for the generalized Langevin dynamics
(\ref{langeq}) as a function of the initial nuclear temperature $T$
at quite small memory time $\tau=2*10^{-23}~s$ (points line) and
at fairly large memory time $\tau=8*10^{-23}~s$ (solid line).}
\label{Fig1}
\end{figure}

We see that with the increase of the initial temperature the role
of the non-Markovian features of the macroscopic dynamics (\ref{langeq})
grows.

\section{Summary}
\label{sum}

In the paper, we have considered the response of a complex quantum system
on time variatios of a single macroscopic coordinate $q$.
The driven dynamics has been studied in terms of the time evolution of the
adiabatic occupancies (\ref{adiabbas}) of the system's Hamilton operator
$\hat{H}(q)$. In the limit of weak coupling of the macroscopic coordinate
to the quantum system (\ref{wcr}), we have obtained diffusion--like
equations of motion for the adiabatic occupancies, see Eq.~(\ref{rnonm}).
Within the random matrix model~(\ref{HH})--(\ref{RGSE}), the time features
of the quantum mechanical diffusion~(\ref{rnonm}) have been investigated.

Thus, we have found the normal regime of the diffusion (\ref{dif0})
with the diffusion coefficient quadratically proportional
to the driving velocity $\dot{q}$. This regime is realized for
sufficiently spread $\Gamma$ energy distribution of the coupling
matrix elements $\partial \hat{H}/\partial q$ (\ref{HH}), when the
characteristic time $\hbar/\Gamma$
is the shortest time scale in the system. In the opposite limit of
quite small values of the spreading width $\Gamma$, we get fully
ballistic regime (\ref{dif8}) of the quantum driven dynamics
(\ref{rnonm}). Here, the energy transport in a space of the adiabatic
occupancies undergoes as a process of a wave propagation. At moderate
values of $\Gamma$, the ballistic and diffusive regimes of the driven
dynamics coexist.

Within the cranking approach~(\ref{dEdt}), we have made an attempt to
measure the macroscopic manifestation of the quantum mechanical diffusion
(\ref{rnonm}). For the first time, we have naturally include into
the standard framework of the cranking approach fluctuative properties
of the macroscopic classical coordinate $q$. We have also established
memory effects in the motion of the macroscopic coordinate appearing
due to the time features of the intrinsic diffusive dynamics.

We used the obtained equation for the macroscopic variable (\ref{eqm1})
to the study of a nuclear escape over a model parabolic fission barrier.
In practise, it was numerically solved the generalized Langevin equation
of motion (\ref{langeq}) for the nuclear shape parameter $q$ and calculated
the fission rate $R_f$ at different values of the memory time $\tau$.
We found that the memory effects in the nuclear fission dynamics
become stronger with the growth of the initial temperature
of a nucleus (see Fig.~1).

\section{Acknowledgements}

The work of S.V.R. on the project “Nuclear collective dynamics for
high temperatures and neutron-proton asymmetries” was supported
(partially) by the Program “Fundamental research in high energy
physics and nuclear physics (international collaboration)” at
the Department of Nuclear Physics and Energy of the National
Academy of Sciences of Ukraine.

\end{document}